\documentclass[12pt,a4paper]{article}

\usepackage{amsfonts}

\usepackage{epsfig}
\usepackage{graphics}

\newcommand{\vol}[1]{{\bf #1}}

\newcommand{\ttitle}[1]{{\it #1}}
\newcommand{\tpretitle}[1]{}
\newcommand{\arttitle}[1]{}
\newcommand{\inproctitle}[1]{``#1''}

\newcommand{\tnote}[1]{}
\newcommand{\tcomment}[1]{}
\newcommand{\tnotpre}[1]{#1}
\newcommand{\tpre}[1]{}
\newcommand{\tprenote}[1]{}

\newcommand{\href}[2]{#2}
\newcommand{\eprint}[1]{{\tt #1}}

\newcommand{\tsedevelop}[1]{{}}

\renewcommand{\tnotpre}[1]{}
\renewcommand{\tpre}[1]{#1}
\renewcommand{\tprenote}[1]{\footnote{#1}}
\renewcommand{\href}[2]{{#2}{}}
\renewcommand{\eprint}[1]{\href{http://xxx.soton.ac.uk/abs/#1}{{\tt #1}}}
\renewcommand{\vol}[1]{{\bf #1}}

\renewcommand{\ttitle}[1]{{\em #1}}
\renewcommand{\tpretitle}[1]{{\em #1}}

\renewcommand{\inproctitle}[1]{``#1''}
\renewcommand{\arttitle}[1]{{\em #1},}

\def\be{\beta}

\def\de{\delta}

\def\ka{\kappa}
\def\la{\lambda}

\def\La{\Lambda}

\def\Om{\Omega}

\def\l{\left}
\def\r{\right}

\newcommand{\beq}{\begin{equation}\label}
\newcommand{\eeq}{\end{equation}}
\newcommand{\beqa}{\begin{eqnarray}\label}
\newcommand{\eeqa}{\end{eqnarray}}

\begin{document}

\renewcommand{\thefootnote}{\fnsymbol{footnote}}

\tpre{\begin{flushright} {\tt Imperial/TP/99-0/038} \\
\eprint{hep-ph/0008307} \\
10th August 2000 \\
\tsedevelop{ (LaTeX-ed on \today ) }
\end{flushright}
\vspace*{1cm}}

\begin{center}
{\Large\bf  Non-perturbative calculations of a}
\\ {\Large\bf global U(1) theory at finite density and temperature }
\\ {\tpre{ \vspace*{1cm} } }
{\large
T.S.Evans\footnote{email: \href{mailto:T.Evans@ic.ac.uk} {\tt T.Evans@ic.ac.uk}
    \tpre{, WWW: \href{http://euclid.tp.ph.ic.ac.uk/links/time}
    {\tt http://theory.ic.ac.uk/\symbol{126}time} }},
H.F.Jones\footnote{email: \href{mailto:H.F.Jones@ic.ac.uk} {\tt H.F.Jones@ic.ac.uk}}
and D.Winder\footnote{email: \href{mailto:d.winder@ic.ac.uk} {\tt d.winder@ic.ac.uk}
    \tpre{Tel: [+44]-20-7594-7839,
    Fax: [+44]-20-7594-7844.}}
\\[1cm]
}
\href{http://euclid.tp.ph.ic.ac.uk/}
Theoretical Physics, Blackett Laboratory, Imperial College, \\
Prince Consort Road, London, SW7 2BZ,  U.K.
\end{center}

\vspace*{1cm}

\begin{abstract}
We use an optimised hopping parameter expansion for the free energy 
(linear $\delta$-expansion) to study the phase transitions at finite temperature and 
finite charge density in a global U(1) scalar Higgs sector on the lattice at large lattice 
couplings. We are able to plot out phase diagrams in lattice parameter space and find that 
the standard second-order phase transition with temperature at zero chemical potential 
becomes first order as the chemical potential increases.
\end{abstract}

\renewcommand{\thefootnote}{\arabic{footnote}}
\setcounter{footnote}{0}

% *******************************************************
% *******************************************************

% --- Introduction ---
\section{Introduction}\label{intro}

Quantum field theory has had many remarkable successes in the past fifty years
using the standard methods of perturbative expansions. However, there are many
situations where such techniques are not appropriate. The obvious example is QCD,
which has strongly coupled bound states at low momenta. However, phase transitions
at non-zero temperatures and densities, even in models without infrared slavery
like the electroweak model, are also poorly explained by perturbative
methods. It is this latter example which motivates our interest in scalar Higgs fields.
The global U(1) model is also a relatively simple testing ground for techniques at finite
chemical potential and temperature. There are many examples of physical situations
at finite charge densities. Any phase transitions after baryogenesis in the early universe
occur at finite baryon charge density. Heavy ion collisions also occur in the same regime.

Having ruled out a perturbative approach we turn to the various available non-perturbative
techniques. One's first thought is to use Monte Carlo (MC) techniques to tackle the
problem. However, MC methods usually fail when considering models at finite
densities. This is because the Euclidean action, which is used as a statistical
weight for the system, becomes complex, making a simple statistical integration technique
imposible.

To replace the MC approach we need an analytical non-perturbative method.
The method chosen in this paper is an example of a general family
sometimes called linear $\de$-expansions \cite{Jo}. Examples of
these methods have appeared under many names, including optimised pertubation
theory\cite{EIM}, action-variational approach \cite{KM}, variational perturbation
theory \cite{SSS}, method of self-similar approximations \cite{Yu}, screened
perturbation theory \cite{ABS},and the variational cumulant expansion \cite{Wu}.
The method has been applied successfully to\footnote{The papers quoted are not
necessarily the first in their area but are usually
good starting points for examining the literature in greater detail.}
the evaluation of simple integrals \cite{Jo,BuDJ,BeDJ}, solving non-linear differential
equations \cite{BMPS}, quantum mechanics \cite{Jo,DJ,Ca,Ki,Ok},
cosmological slow roll transitions \cite{JPW} and to quantum field theory,
both in the continuum \cite{Ok,SSS} and on the lattice
\cite{Jo,AJ93a,AJ93b,AJP,EJR98a,EJR98b,KM,ZTW,ZL,Wu,Ya,YWZ}.
Since the LDE approach is analytical, we do not have to worry about the presence of a complex
action. The expectation value of all physical observables will turn out to be real.

The work in this paper with the $U(1)$ or $O(2)$ model builds on
that set out in \cite{EIM} for the case of zero temperature and zero chemical
potential. However, here we phrase the model in terms of the field and its
conjugate $(\Phi,\Phi^*)$ rather than working with the real components of the field.
This change is made because the charge operator is diagonal in this representation and so
much easier to deal with\cite{E}.

This paper is also complementary to the work done on the $U(1)$ model using LDE methods in the
continuum at finite temperature and finite chemical potential \cite{JP}.

% --- Linear Delta Expansion ---
\section{The Linear Delta expansion}\label{LDE}

The general format of the LDE method is to take a given expansion, whether it be a
perturbative expansion in the continuum or a cumulant expansion on the lattice, and
to provide an {\it order by order optimisation} of this expansion. It is in this process of
optimisation that non-perturbative information emerges. It is
straightforward, in principle, to expand beyond leading order, unlike other non-perturbative
methods like large-N expansions or mean-field approximations.

The first step in the method is to replace the physical action with an interpolating action
made up of a linear combination of the physical action and a {\it soluble} trial action. This trial
action is characterised by some set of new variational parameters $\{\vec{v}\}$. The particular
choice of trial action is the main decision to be made in implementing the method.
The more general the trial action, and the more terms one includes, the
greater the number of variational parameters required to characterise it. The
interpolating action takes the form
$ S \rightarrow S_{\de} = S_0(\vec{v}) + \de ( S - S_0(\vec{v})) $.
If we set $\de = 1$ in the interpolating action then $S_{\de} = S$. The
new statistical average of an operator $ O $ is

\beq{obs_stat_av}
\l\langle O \r\rangle_{\delta} =
\frac{1}{Z_{\de}} \int {\cal D} \Phi \; O \; e^{-S_0} e^{\de(S-S_0)}
\eeq

We now perform our chosen expansion technique on this new action. In this paper we use an
expansion for the free energy on the lattice in terms of cumulant averages (sometimes
known as a linked cluster expansion). However, we expand in the unphysical
parameter $\de$ instead of the coupling or the hopping parameter. The new expansion, as
compared to the original, has additional terms at each order which depend upon the variational
parameters. When we truncate the expansion at a given power of $\delta$ and then
impose $\delta = 1$ there remains a residual dependence on these varational parameters. So
the LDE up to $\de^R$ for the expectation value of some operator $\hat{\mathcal{O}}$ is
\beq{stat_av_LDE}
\l\langle \mathcal{O} \r\rangle_R  = \l\langle \mathcal{O} \r\rangle_R \l( \vec{v} \r)  :=
\lim_{\de \rightarrow 1} \l\{ \l\langle \mathcal{O} \r\rangle_{\de}
\l( \vec{v} \r) \r\}_R
\eeq
where the $\{ \}_R$ bracket signifies the truncation of the power series to order $\de^R$.
It is how we choose to fix the variational parameters, {\it order by order} in
the expansion, which introduces the main fully non-perturbative effects. There are at least two
apparently different approaches to this final part of the method listed in the literature.
The first is to demand that the series converge as fast as possible by minimising the highest
order term with respect to the lower order terms \cite{KM} - the principle of fastest convergence.
However, we shall take the second main approach, often called the principle of minimal
sensitivity (PMS)\cite{PMS}. This is discussed in \cite{Jo,AJ93a,AJ93b,AJP} and used in
standard partical physics to minimise dependence of perturbative results on the renormalisation
scheme choice\cite{PDB}.
For the PMS case the parameters, $\{\vec{v}\}$, are set {\it at each order} in the
expansion by demanding that the variation of some physical observable be zero.
It is useful to be aware of the general statement that a derivative of (\ref{stat_av_LDE}) with
respect to some arbitary parameter $x$ gives
\beqa{PMS_div}
\frac{\partial \l\langle \mathcal{O} \r\rangle_R}{\partial x} =
                            \l\langle \frac{\partial \mathcal{O}}{\partial x} \r\rangle_R
&-&
\lim_{\de \rightarrow 1} \l[ \l\langle \frac{\partial S_0}{\partial x} \mathcal{O} \r\rangle_{\de} -
                             \l\langle \frac{\partial S_0}{\partial x} \r\rangle_{\de}
                             \l\langle \mathcal{O} \r\rangle_{\de}
                         \r]_{\rm R^{\it th} \, term} \nonumber\\
&-&
\lim_{\de \rightarrow 1} \l\{ \l\langle \frac{\partial S}{\partial x} \mathcal{O} \r\rangle_{\de} -
                              \l\langle \frac{\partial S}{\partial x} \r\rangle_{\de}
                              \l\langle \mathcal{O} \r\rangle_{\de}
                         \r\}_{R-1}
\eeqa
The first term is just an LDE up to order $R$, the second is just the last term in an LDE expansion
up to order $R$ and the final term is an LDE up to order $R-1$. This means that the PMS condition
on a variational parameter is
\beqa{PMS_cond}
\frac{\partial \l\langle \mathcal{O} \r\rangle_R}{\partial v} = 0
\Longrightarrow
\lim_{\de \rightarrow 1}
\l[ \l\langle \frac{\partial S_0}{\partial v} \mathcal{O} \r\rangle_{\de} -
    \l\langle \frac{\partial S_0}{\partial x} \r\rangle_{\de}
    \l\langle \mathcal{O} \r\rangle_{\de}
                         \r]_{\rm R^{th} term} = 0
\eeqa
In the special case of the free energy, defined by $F = -  (1 / N) \ln Z$, we have instead
\beq{div_F}
\frac{\partial F_R}{\partial x} =
 \frac{1}{N} \lim_{\de \rightarrow 1} \l[\l\langle \frac{\partial S_0}{\partial x} \r\rangle_{\de}
                         \r]_{\rm R^{th} term}
+
 \frac{1}{N} \lim_{\de \rightarrow 1} \l\{\l\langle \frac{\partial S}{\partial x} \r\rangle_{\de}
                         \r\}_{R-1}
\eeq

\beqa{PMS_cond_F}
\frac{\partial F_R}{\partial v} = 0
\Longrightarrow
\lim_{\de \rightarrow 1}
\l[ \l\langle \frac{\partial S_0}{\partial v} \r\rangle_{\de} \r]_{\rm R^{th} term} = 0
\eeqa
In general the PMS condition is only a condition on the final term in the expansion. In this
way one sees that the PMS and principle of fastest convergence are quite closely related.

The variational parameters are set order by order by (\ref{PMS_cond}) or (\ref{PMS_cond_F})
and then substituted back into the expression for the physical observable to obtain the
LDE estimate for its value. In quantum mechanical models
at least, the absolute convergence of this LDE estimate to the true solution can be
proven\cite{DJ}.

We have set out the two overall approaches to fixing the varational parameters. However, the
choice of what physical observable one chooses for the PMS procedure vastly increases the
possible options. One can for example always apply the PMS techniques to an LDE for the
particular observable one is trying to measure; this could be called a true PMS approach. On
the other hand one could choose to extremise an LDE for one particular physical observable, say
the free energy, and then apply the values of the variational parameters gleaned therefrom to
another LDE for the particular physical observable one is interested in. These two options are
discussedin \cite{EIM} for the $O(2)$ scalar model and the second option is preferred.

We have set out the two overall approaches to fixing the
variational parameters. However, there is a further degree of
freedom in the choice of which physical observable to apply the
PMS procedure to. The strict LDE approach would be to apply the PMS
procedure to the expansion for the observable one is interested in.
However, in some cases (see, for example, Ref.~4 for the
$O(2)$ scalar model) there may be no PMS points in that expansion.
The alternative is to fix the variational parameters by applying
the PMS procedure to the expansion for one particular, privileged
observable, e.g. the free energy, and then use those parameters in 
the expansions for all other observables.

In this paper we shall take what is effectively the second approach, but instead
of calculating two expansions, one for the free energy and one for the physical parameter we
are trying to evaluate, we will evaluate the free energy using an LDE approach with the PMS
criterion and then calculate all the other required physical observables by {\it numerical}
differentiation with respect to the physical parameters of the free energy.

%U(1) Model on the lattice
\section{The U(1) complex scalar model on the lattice}\label{lattice}

We work with a Euclidean time formulation of the $U(1)$ field theory. In the continuum the
global U(1) model at finite density has a conserved charge of the
form $ {\cal Q}= i ( \Phi^* \Pi - \Phi \Pi^* )$. The presence of a finite charge density
at finite temperature gives rise to an effective action\cite{E}
\beqa{S_eff}
S_{\rm eff} = - \int_{0}^{\be} dt \int d^3x \l[
(\nabla_\mu \Phi)^* (\nabla^\mu \Phi)
+ ( \mu_0^2 - m_0^2) \Phi^* \Phi - \la_0 (\Phi^* \Phi)^2
\r.
\nonumber\\
\l.
- i \mu_0 ( \Phi^* \dot{\Phi} - \Phi \dot{\Phi^*} )
+ J_0^* \Phi + J_0 \Phi^* \r]
\eeqa
The temperature appears as a finite boundary condition in the Euclidean
time direction, where $ N_t a_t = 1/T $. The integer $N_t$ is the number of lattice links in the
temporal direction and $a_t$ is the temporal lattice spacing. In contrast the chemical
potential is present in the main body of the Lagrangian. In particular we have a slightly
surprising $\mu_0^2$ term which arises from integrating out the conjugate fields when
constructing the effective action. This term can cause symmetry breaking with
positive $m_0^2$, even for free fields, when the transition occurs at $\mu = m_L$. In a
Euclidean formulation the effective action is complex. This complexification caused by
the introduction of a chemical potential is generic.
In addition, the presence of the chemical potential means that the action is not invariant
under $t \rightarrow -t$ (Osterwalder-Schrader reflection)\cite{MM}.

To regularise the UV infinities of quantum field theory we introduce a lattice
formalism and replace Euclidean spacetime with a hypercubic 4d lattice.
The lattice points are denoted by $n \equiv (n_4, \bf{n})$. With a standard redefinition
of the physical parameters and fields we get
\beqa{S_discrete}
S = \sum_n
\l[
-\ka_s \sum_{i} \l[ \Phi_{n_4,\bf{n}}^*
\Phi_{n_4,\bf{n}+\bf{e}_i} + \Phi_{n_4,\bf{n}} \Phi_{n_4,\bf{n}+\bf{e}_i}^* \r]
- \ka_t (1 + \mu_L) \Phi_{n_4,\bf{n}}^* \Phi_{n_4+1,\bf{n}}
\r.
\nonumber\\
\l.
- \ka_t (1 - \mu_L) \Phi_{n_4,\bf{n}} \Phi_{n_4+1,\bf{n}}^*
+ (m_L^2 - \mu_L^2) \Phi_n^* \Phi_n
+ \la_L ( \Phi_n^* \Phi_n )^2 - J_L^{*} \Phi_n - J_L \Phi_n^* \r]
\eeqa
where $ J_L = ( J_1 +  i J_2 ) / \sqrt{2} $. The physical lattice parameters are
all dimensionless and so measured in units of the lattice cutoff. The derivative
terms have become nearest-neighbour interactions.
We also introduce {\it separate} hopping parameters, $\ka_s$ and $\ka_t = 1$, for
the spatial and temporal nearest-neighbour interaction. The $\ka_t$ has a physical
value of unity but we introduce it as an additional arbitary parameter to allow us to
take derivatives with respect to it later on. The different hopping parameters allow for a
different effective lattice spacing in the Euclidean time direction as compared to the
spatial direction, thus
allowing one to indirectly vary the lattice temperature {\it continuously}, while keeping
the number of  links in the temporal direction constant, by varing $\ka_s$. The detail of
this process is discussed in section \ref{T}. To keep calculations relatively easy we will
keep the Euclidean temporal direction {\it two} links in extent, $N_t = 2$. This means that
we will be working with only the first three Matsubara modes and so effectively at high
temperatures.

As the physical action stands, when we set $J_L = 0$ we have automatically, via symmetry
arguments, that $ \l\langle \Phi \r\rangle = 0 $, so the symmetry is unbroken.
This remains the case whether one is considering a perturbative expansion in the continuum
or a hopping parameter expansion on the lattice.

If the nearest-neighbour terms were to go to zero then the remaining action, containing
only {\it ultralocal} terms, would be exacly soluble. On the lattice we make an
expansion around this solution in powers of the nearest-neighbour terms,
not in powers of the coupling. This is often called a hopping parameter expansion. It is
the hopping parameter expansion for the free energy, often given the special title of a
linked cluster expansion \cite{Wo}, which we will optimise in this paper using LDE techniques.

% --- LDE applied to U(1) model on the lattice ---
\section{LDE applied to U(1) model on the lattice}\label{LDElattice}

The groundwork for the choice of trial action is set out in \cite{EIM}. Following the
lead of this paper we shall include a source term, a quadratic term and a quartic term.
Unfortunately, because we are dealing with an action in terms of the
field and its conjugate rather than its real components, defined
by $\Phi = \l( \phi + i \psi \r) / \sqrt{2}$, we will have to change the notation
slightly\footnote{To compare the two papers set $k_{+} =  -\Omega^2$ , $k_{-} = 0$
, $j_{\phi} = j_1$ and $j_{\psi} = j_2$. The physical parameters are also defined slightly
differently: $\la_{EJW} = 4 \la_{EIM}$ and $m^2_{EJW} = 2 m^2_{EIM}$. }.
\beqa{S_0}
S_0 = \sum_n
\l[  (\Omega^2 - \mu_L^2) \Phi^{n *} \Phi^n
+ \la_L ( \Phi^{n *} \Phi^n )^2 - j^{*} \Phi^n - j \Phi^{n *} \r]
\eeqa
where $j = \l( j_1 + i j_2 \r) / \sqrt{2}$. With this choice of $S_0$ we obtain
an $S_{\delta}$ which we organise as follows:

\beqa{S_delta}
S_{\delta} &=& S_0 + \delta S_1 - \delta S_{int} = S_U(\delta) - \delta S_I\nonumber\\
S_1 &=&  \sum_n
\l[
 (m_L^2 - \Omega^2) \Phi_n^* \Phi_n - (J_L^{*} - j^{*}) \Phi_n - (J_L - j) \Phi_n^* \r]
\nonumber\\
S_{int} &=& \sum_n \l[
\kappa_s \sum_{i} \l[ \Phi_{n_4,\bf{n}}^*\Phi_{n_4,\bf{n}+\bf{e}_i} +
\Phi_{n_4,\bf{n}} \Phi_{n_4,\bf{n}+\bf{e}_i}^* \r]
+\kappa_t (1 + \mu_L) \Phi_{n_4,\bf{n}}^* \Phi_{n_4+1,\bf{n}}
\r.
\nonumber\\
&&\l.
+\kappa_t (1 - \mu_L) \Phi_{n_4,\bf{n}} \Phi_{n_4+1,\bf{n}}^* \r]
\eeqa

Although the $S_U$ is still $\de$ dependent it is {\it ultralocal}, i.e. it contains
no interaction terms between lattice points. We now perform a diagrammatic expansion of
the $S_I$ term identical to the standard hopping parameter {\it link} expansion. As it stands
this is not strictly a $\de$ expansion, as there is some residual $\de$ dependence in the $S_U$
action. However, this can be dealt with separately later on in the calculation. The
expansion is best represented graphically, where we define the spatial and temporal links as

\setlength{\unitlength}{0.1cm}

\beqa{s_and_t_links}
L_s &:=& \ka_s \sum_{n, i} \l[ \Phi_{n_4,\bf{n}}^* \Phi_{n_4,\bf{n}+\bf{e}_i} +
\Phi_{n_4,\bf{n}} \Phi_{n_4,\bf{n}+\bf{e}_i}^* \r] :=
\sum_{n, i}
\begin{picture}(10,1)
\put(1,1){\circle*{0.5}}
\put(1,1){\line(1,0){8}}
\put(9,1){\circle*{0.5}}
\end{picture} \nonumber\\
L_t &:=& \sum_n \l[ \ka_{t} \r (1 + \mu_L \l) \Phi_{n_4,\bf{n}}^* \Phi_{n_4+1,\bf{n}}
+ \ka_{t} \r (1 - \mu_L \l) \Phi_{n_4,\bf{n}} \Phi_{n_4+1,\bf{n}}^* \r] :=
\sum_n
\begin{picture}(4,8)
\thicklines
\put(2,-3){\circle*{0.5}}
\put(2,-3){\vector(0,1){8}}
\put(2,5){\circle*{0.5}}
\end{picture}
\nonumber\\
\rightarrow S_{int} &=& \sum_n \l[
\begin{picture}(4,8)
\thicklines
\put(2,-3){\circle*{0.5}}
\put(2,-3){\vector(0,1){8}}
\put(2,5){\circle*{0.5}}
\end{picture}
+\sum_{i}
\begin{picture}(10,1)
\put(1,1){\circle*{0.5}}
\put(1,1){\line(1,0){8}}
\put(9,1){\circle*{0.5}}
\end{picture} \r]
\eeqa

Thick vertical lines are in the temporal direction, and thin horizontal or diagonal lines
represent spatial directions. An arrow is added to the temporal link because, due to the presence
of a non-zero chemical potential, $L_t$ is {\it not symmetric} under interchange of
initial and final lattice points, this is due to the lack of Osterwalder-Schrader symmetry
in the starting action (\ref{S_eff}).
The arrow can be thought of as representing a net flow of charge in the imaginary `time' direction.

To extract the thermodynamical information we calculate the free energy per lattice site in
terms of {\it cumulant} averages. To do this we first construct statistical averages with
respect to the $S_U$ action (\ref{U_stat_av}); these are given a subscript $U$.
\setlength{\unitlength}{0.1cm}
\beqa{U_stat_av}
\l\langle \mathcal{O} \r\rangle_{U} &:=&
\frac{1}{Z_U} \int {\cal D} \Phi \; \mathcal{O} \; e^{-S_U} \nonumber\\
\Rightarrow Z &=& Z_U(\delta) \l\langle e^{\delta S_I} \r\rangle_U = Z_U(\delta)
\sum_{j=0}^{R}
\l[ \frac{\delta^j}{j!} \l\langle \l(
 \sum_n \l[
\begin{picture}(4,8)
\thicklines
\put(2,-3){\circle*{0.5}}
\put(2,-3){\vector(0,1){8}}
\put(2,5){\circle*{0.5}}
\end{picture}
+\sum_{i}
\begin{picture}(10,1)
\put(1,1){\circle*{0.5}}
\put(1,1){\line(1,0){8}}
\put(9,1){\circle*{0.5}}
\end{picture} \r]
\r)^j \r\rangle_U \r]
\eeqa
Then cumulant averages, denoted by a subscript $C$, are constructed from the $S_U$ statistical averages
according to
\beq{cumulant}
\sum_{j=1}^{\infty} \frac{1}{j!} \l\langle O^j \r\rangle_C :=
\ln \l[ \l\langle e^O \r\rangle_U
\r]
\eeq
We thus obtain the following expression for the free energy
\setlength{\unitlength}{0.1cm}
\beq{F}
F = - \frac{1}{N} \ln Z_U
- \frac{1}{N} \sum_{j=1}^{\infty} \frac{\delta^j}{j!} \l\langle
\l(
\sum_{n}
\begin{picture}(4,8)
\thicklines
\put(2,-3){\circle*{0.5}}
\put(2,-3){\vector(0,1){8}}
\put(2,5){\circle*{0.5}}
\end{picture}
+\sum_{n,i}
\begin{picture}(10,1)
\put(1,1){\circle*{0.5}}
\put(1,1){\line(1,0){8}}
\put(9,1){\circle*{0.5}}
\end{picture} \r)^j
\r\rangle_C
\eeq
Because this is in terms of cumulant averages we only have to consider connected diagrams.
As an example, the $j=2$ term is:
\setlength{\unitlength}{0.05cm}
\beqa{F_term}
\l\langle \l(
\sum_{n}
\begin{picture}(4,8)
\thicklines
\put(2,-3){\circle*{0.5}}
\put(2,-3){\vector(0,1){8}}
\put(2,5){\circle*{0.5}}
\end{picture}
+\sum_{n,i}
\begin{picture}(10,1)
\put(1,1){\circle*{0.5}}
\put(1,1){\line(1,0){8}}
\put(9,1){\circle*{0.5}}
\end{picture} \r)^2 \r\rangle_C &=&
N \l\langle
\begin{picture}(4,8)
\thicklines
\put(2,-3){\circle*{0.5}}
\put(1,-3){\vector(0,1){8}}
\put(3,-3){\vector(0,1){8}}
\put(2,5){\circle*{0.5}}
\end{picture} \r\rangle_C
+ 2 N \l\langle
\begin{picture}(4,16)
\thicklines
\put(2,-6){\circle*{0.5}}
\put(2,-6){\vector(0,1){8}}
\put(2,2){\circle*{0.5}}
\put(2,2){\vector(0,1){8}}
\put(2,10){\circle*{0.5}}
\end{picture} \r\rangle_C
+ 2 N(d-1) \l\langle
\begin{picture}(12,8)
\thicklines
\put(2,-3){\circle*{0.5}}
\put(2,-3){\vector(0,1){8}}
\put(2,5){\circle*{0.5}}
\thinlines
\put(2,-3){\line(1,0){8}}
\put(10,-3){\circle*{0.5}}
\end{picture} \r\rangle_C \nonumber\\
+ 2 N(d-1) \l\langle
\begin{picture}(12,8)
\thicklines
\put(2,-3){\circle*{0.5}}
\put(2,-3){\vector(0,1){8}}
\put(2,5){\circle*{0.5}}
\thinlines
\put(2,5){\line(1,0){8}}
\put(10,5){\circle*{0.5}}
\end{picture} \r\rangle_C
&+& N(d-1)\l\langle
\begin{picture}(10,1)
\put(1,2){\circle*{0.5}}
\put(1,1){\line(1,0){8}}
\put(1,3){\line(1,0){8}}
\put(9,2){\circle*{0.5}}
\end{picture} \r\rangle_C
+ N(d-1)\l\langle
\begin{picture}(18,1)
\put(1,1){\circle*{0.5}}
\put(1,1){\line(1,0){8}}
\put(9,1){\circle*{0.5}}
\put(9,1){\line(1,0){8}}
\put(17,1){\circle*{0.5}}
\end{picture} \r\rangle_C
\eeqa
where $d$ is the number of spacetime dimensions. The multiplicities are calculated by hand
for each of the distinct diagrams. As one goes up in order the number of diagrams in the cumulant
expansion increases rapidly. The hopping parameter expansion has been evaluated at least up
to $14^{th}$ order for an $O(N)$ scalar field theory at zero temperature and zero
chemical potential \cite{LW1,LW2,HR}. There have also been some calculations at finite temperature
\cite{R1,R2}.

We take the hopping parameter expansion for $F$ up to third order, calculating the diagrams
by hand. The full list of diagrams and multiplicities used is given in Appendix \ref{diagrams}.
Having performed the standard linked cluster expansion to a given order, with the additional
diagrammatic complications introduced by the presence of the chemical potential, we need to
convert the expansion to one in terms of powers of $\delta$. To do this we have to Taylor
expand the cumulant averages to include the residual $\delta$ dependence caused by the fact
that $S_U = S_U(\delta)$. This will introduce additional optimisable terms at each order
in the expansion. This method of calculating the LDE is much simpler than including all
the $\de$ dependent terms in $S_{int}$ at the start, which requires many additional diagrams.

Not all the diagrams in the hopping parameter expansion `feel' the temperature, as they are
not periodic across the Euclidean time direction. However, when we apply the variational
part of the LDE approach all the diagrams effects are combined and so the temperature is
felt indirectly by the whole of the expansion.

The presence of a $j^* \Phi$ term in the trial action, where the $j$ value is
fixed {\it variationally}, allows for the possibility that $\l\langle \Phi \r\rangle \not= 0$
even when $J_L = 0$. This means that we can examine the symmetry breaking of the model using
the optimised LDE version, unlike the standard free energy expansion,  which is fixed
on one side of the phase transition by the symmetry of the physical action.

% --- example diagram calculation ---
\section{Example diagram calculation}\label{example}

The statistical and cumulant averages result in an expression made up of known
parameters and a set of integrals involving the $L_U$ Lagrangian. These integrals in turn, when
one performs the Taylor expansion in $\de$, become integrals with respect to $L_0$
because $L_U = L_0 + \de L_1$. At this point we should define the general set of integrals

\beqa{integrals}
J_{m n}(\de) &:=& \int d \phi d \phi^* (\phi)^m (\phi^*)^n e^{-L_U(\de)} \qquad
K_{m n}(\de) := \int d \phi_2 d \phi_1 (\phi_1)^m (\phi_2)^n e^{-L_U(\de)} \qquad
\nonumber\\
B_{m n} &:=& \int d \phi_2 d \phi_1 (\phi_1)^m (\phi_2)^n e^{-L_0}
\nonumber\\
&& \hat{J}_{m n}(\de) := \frac{J_{m n}(\de)}{J_{0 0}(\de)} \qquad
\hat{K}_{m n}(\de) := \frac{K_{m n}(\de)}{K_{0 0}(\de)} \qquad
\hat{B}_{m n} := \frac{B_{m n}}{B_{0 0}}
\eeqa
In terms of these definitions,

\beq{Z_U_result}
\ln Z_U = N \ln J_{0 0}
\eeq

All the diagrams we need to calculate are expressible in terms of these
integrals. Consider the very simple example of the cumulant average of the single
spatial link. The cumulant as defined in (\ref{cumulant}) is expandable in terms
of statistical averages with respect to the $S_U$ action (\ref{U_stat_av}). We get

\setlength{\unitlength}{0.1cm}

\beqa{single_s_link1}
&&
\l\langle
\begin{picture}(10,1)
\put(1,1){\circle*{0.5}}
\put(1,1){\line(1,0){8}}
\put(9,1){\circle*{0.5}}
\end{picture}
\r\rangle_C =
\l\langle
\begin{picture}(10,1)
\put(1,1){\circle*{0.5}}
\put(1,1){\line(1,0){8}}
\put(9,1){\circle*{0.5}}
\end{picture}
\r\rangle_U
\nonumber\\
&&=
\ka_s \l[ \l\langle \phi_{n_4, \bf{n}}^* \phi_{n_4,\bf{n}+\bf{e}_i} \r\rangle_U +
           \l\langle \phi_{n_4,\bf{n}} \phi_{n_4,\bf{n}+\bf{e}_i}^* \r\rangle_U \r]
=
\ka_s \l[ \l\langle \phi_a^* \phi_b \r\rangle_U +
             \l\langle \phi_a \phi_b^* \r\rangle_U \r]
\nonumber\\
&&=\ka_s \l[  \frac{\int \l( \prod_m d \phi^m  d \phi^{* m} \r)
                         \; \phi_a^* \phi_b \; \exp \l\{- \sum_n L_U^n(\de)\r\} }
                    {\int \l( \prod_m d \phi_m d \phi_m^* \r)
                                                 \exp \l\{- \sum_n L_U^n(\de)\r\} }
\r. \nonumber\\ && \qquad \qquad \qquad \qquad \qquad  \l.
               + \frac{\int \l( \prod_m d \phi_m  d \phi_m^* \r)
                         \; \phi_a \phi_b^* \; \exp \l\{- \sum_n L_U^n(\de)\r\} }
                    {\int \l( \prod_m d \phi_m d \phi_m^* \r)
                                                 \exp \l\{- \sum_n L_U^n(\de)\r\} }
         \r]
\eeqa

As $L_U$ is ultralocal all the integrals separate from each other. It was to ensure
this fact that we imposed ultralocality on our choice of trial action, $S_0$. The
integrals over variables other than $\phi_a$ and $\phi_b$ cancel between the numerator and
denominator. Therefore in both terms in (\ref{single_s_link1}) we will just be left with the product
of two integrals. Using the identities defined in (\ref{integrals}) we get

\setlength{\unitlength}{0.1cm}

\beqa{single_s_link2}
\l\langle
\begin{picture}(10,1)
\put(1,1){\circle*{0.5}}
\put(1,1){\line(1,0){8}}
\put(9,1){\circle*{0.5}}
\end{picture}
\r\rangle_C
=
&&\ka_S \l[ \l\langle \phi^{* a} \r\rangle_U
              \l\langle \phi^{b} \r\rangle_U +
              \l\langle \phi^{a} \r\rangle_U
              \l\langle \phi^{* b} \r\rangle_U \r]
= 2 \ka_S \l\langle \phi^{a} \r\rangle_U
    \l\langle \phi^{* a} \r\rangle_U \nonumber\\
&&= 2 \ka_S \frac{ J_{1 0}(\de) J_{0 1}(\de) }{ J_{0 0}(\de)^2 }
=  2 \ka_S \hat{J}_{1 0}(\de) \hat{J}_{0 1}(\de)
\eeqa

All the cumulants obtained from the expansion in (\ref{F}) are similarly expressible
in terms of some combination of our integrals. The final step in evaluating the expansion
of $F$ to a given explicit order in $\de$ is to Taylor expand the $\hat{J}$ integrals. We
first express the $\hat{J}$ integrals as $\hat{K}$ integrals. Then we Taylor expand the later
using the following recursion relation for the derivative:

\beqa{diff_K}
\frac{ \partial \hat{K}_{m n}}{\partial \de} &=&
- \frac{1}{2} (m_L^2 - \Omega^2) \hat{K}_{m+2 \; n}
- \frac{1}{2} (m_L^2 - \Omega^2) \hat{K}_{m \; n+2}
+ (J_1 - j_1) \hat{K}_{m+1 \; n}  \nonumber\\
&&+ (J_2 - j_2) \hat{K}_{m \; n+1} - \hat{K}_{m n} \l[ - \frac{1}{2} (m_L^2 - \Omega^2) \hat{K}_{2 0}
- \frac{1}{2} (m_L^2 - \Omega^2) \hat{K}_{0 2}
 \r. \nonumber\\ &&\l.
+ (J_1 - j_1) \hat{K}_{1 0} + (J_2 - j_2) \hat{K}_{0 1} \r]
\eeqa

Noting that $ \l. \hat{K}_{m n} \r|_{\de=0} = \hat{B}_{m n} $ we have all the required information to
evaluate the Taylor expansion to any order.  For example, the single spatial link can now be
calculated using

\setlength{\unitlength}{0.1cm}

\begin{equation}
\hat{K}_{1 0}(\de) =\hat{B}_{1 0} + \de \l. \frac{\partial \hat{K}_{1 0}}{\partial \de}
\r|_{\de=0}
+ \cdots \qquad
\hat{K}_{0 1}(\de) = \hat{B}_{0 1} + \de \l. \frac{\partial \hat{K}_{0 1}}{\partial \de}
\r|_{\de=0}
+ \cdots \nonumber
\end{equation}
as
\beq{single_s_link3}
\l\langle
\begin{picture}(10,1)
\put(1,1){\circle*{0.5}}
\put(1,1){\line(1,0){8}}
\put(9,1){\circle*{0.5}}
\end{picture}
\r\rangle_C = 2 \ka_S \l[ \hat{B}_{1 0}^2 + \hat{B}_{0 1}^2 \r]
+ 4 \ka_S \de \l[ \hat{B}_{1 0} \l. \frac{\partial \hat{K}_{1 0}}{\partial \de} \r|_{\de=0}
                 +   \hat{B}_{0 1} \l. \frac{\partial \hat{K}_{0 1}}{\partial \de} \r|_{\de=0}
              \r]
\eeq

All the other cumulant averages, initially evaluated in terms of $\hat{J}$ integrals, and
the $\ln J_{00}$ term, can be similarly expressed as a power series in $\de$ in terms of
the $\hat{B}$ integrals.
As one increases the order of the $\de$-expansion the number
of diagrams soon becomes too large to deal with by hand. At order $3$, for example, there are
roughly $20$ diagrams (see Appendix \ref{diagrams}). We therefore use an algebraic
manipulation program which can handle the repetitious task of calculating
all the Taylor expansions and rearranging the overall expansion in explicit powers of $\delta$.

%----- Minimisation----
\section{Minimisation}\label{mini}

\begin{figure}[hp]
\begin{center}
\scalebox{0.93}{\includegraphics*[2.4cm,7cm][19.5cm,27cm]{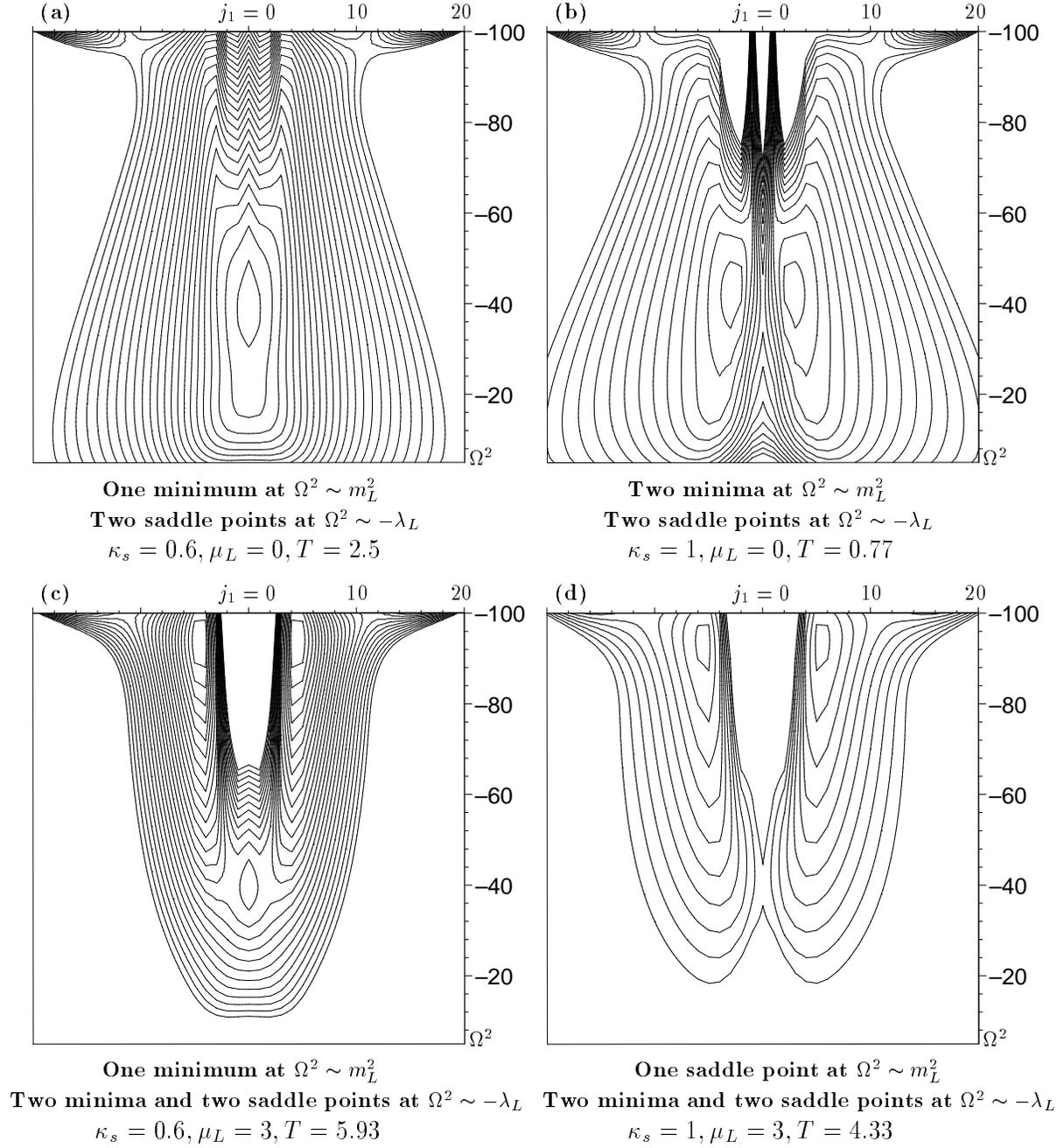}}
\caption{Various minimisation plots.
In all graphs $ m_L^2 = -40 , \ka_t = 1, \la_L = 100, J_1 = J_2 = 0$.}
\label{f_mini}
\end{center}
\end{figure}

As the expression for $F$ given in (\ref{F}) stands, each order in the expansion
has residual dependence on the variational parameters $\l\{ \Omega^2, j_1, j_2 \r\}$, as well
as the expected dependence on the physical
parameters $\l\{ m_L^2, \ka_s, \ka_t, \mu_L, \la_L, J_1, J_2 \r\}$. We will
fix the variational parameters using the PMS condition set out in (\ref{PMS_cond_F}).
But first we note that we can immediately set $j_2 = 0$, which can be justified
using symmetry considerations \cite{EIM}. Having made this choice we are reduced to a
two-dimensional variational parameter space. Minimisation of the free energy in this space
is best represented as a two-dimensional contour plot. The phase transitions in the physical
parameter space are actually mirrored in the variational parameter space in terms of the
number of minima present. This behaviour has already been noted in more limited terms in \cite{EIM}.
As a result of this there are a few generic pictures for the minimisation depending on the values of
the physical parameters. The minimisation curves are plotted in Figure~\ref{f_mini}.

We include the lattice temperature with these plots. See section {\ref{T}} for the derivation
of these temperatures.
Going from Fig.~\ref{f_mini}(a) to~\ref{f_mini}(b) we {\it increase} $\ka_s$ and
so {\it decrease} the lattice temperature. In so doing we move from one minimum
to two minima in the variational parameter space and at the same time undergo a second order phase
transition in the physical parameter space. Similarly, going from Fig.~\ref{f_mini}(a)
to~\ref{f_mini}(c) we increase $\mu_L$ (also indirectly increase $T$) and in so doing move from one
minimum to three minima in the variational parameter space. At around the same time in
the physical parameter space we undergo a first-order phase transition. The order of
the `phase transition' in variational parameter space matches the order of the true
phase transition in the physical parameter space.

In the unbroken physical phase the minima will be found at the point:
\beqa{mini_unbroken}
\Om^2 = m^2, \qquad j_1 = 0
\eeqa
This is discussed further for the zero temperature and zero chemical potential case
in \cite{EIM}. As noted in that paper, because the minimisation space for the free energy
reflects the physical parameter phase space, the value of the variational parameters
at the minimum is often sufficient information to find the phase of the system, without a full
calculation of $\l\langle \Phi \r\rangle$.  However, in the case of a first-order
phase transition more care is needed, as there will be both `broken' and `unbroken' minima,
as in Fig.~\ref{f_mini}(c). One chooses between them by taking the overall minimum. As an aside, note
that minimising the free energy is equivalent to maximising the entropy, which gives us
a further physical justification for our particular PMS approach.

By tracking the value of $F$ at all the minima, and taking special care when we have a
first-order transition, we can build up a set of $F$ values for any range of physical parameters.
From the free energy data we can then evaluate many other correlators by taking numerical
derivatives with respect to the physical parameters.

Of course the method is always limited by the fact that it requires the presence of
a minimum. When the minima in $F$ disappear the method breaks down. However, the
method is sustained far enough on either side of the phase transition for a large
enough range of physical parameters to allow it to be used to plot out reasonable phase
diagrams.

Having minimised the free energy we can now take {\it numerical} derivatives of $F$ to
evaluate particular physical observables. This will allow us to build up phase portraits in
the $\l\{T,\mu_L \r\}$ plane for any choice of the other physical parameters.

%------------------------------------------RESULTS----------------------------------------
\section{Numerical Results}

% --- Phase transition using Ks ---
\subsection{Phase transition using $\ka_s$}\label{phase_Ks}

As a first example we shall look at evaluating $ \l\langle \Phi_1 \r\rangle $ as a
function of the hopping parameter, $\ka_s$, to look for a phase transition. To do this
we evaluate the estimated $F$ value, by tracking the minima across our 2D variational
parameter space, at two nearby $J_1$ values around zero to allow us to evaluate a
numerical derivative with respect to $J_1$. This is related to
the $ \l\langle \Phi_1 \r\rangle $ value using (\ref{div_F}) to give

\beq{phi1}
\l. \frac{\partial F}{\partial J_1} \r|_{J_1 = 0} = - \frac{1}{N} \l\langle
\sum_n \Phi_{1 n} \r\rangle =
- \l\langle \Phi_1 \r\rangle
\eeq

Therefore by taking numerical derivatives we can look for symmetry breaking,
where $ \l\langle \Phi_1 \r\rangle \not= 0$. For simplicity we consider
the $\mu_L = 0$ curves. In this case the variational parameter space tracks between
Fig.~\ref{f_mini}(a) and Fig.~\ref{f_mini}(b).

\begin{figure}[hp]
\begin{center}
\scalebox{1}{\includegraphics*[6.5cm,18.2cm][15cm,27cm]{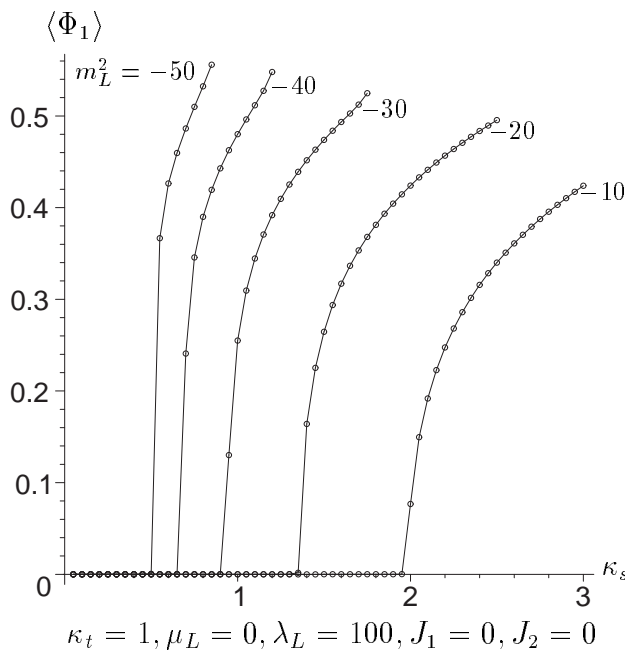}}
\caption{Phase transition with $\ka_s$}
\label{f_Ks}
\end{center}
\end{figure}

We have a second order phase transition to a {\it broken} symmetry sector as we increase
the $\ka_s$ value, and so decrease the lattice temperature. This transition happens at lower
and lower temperatures as one increases the lattice mass.

% --- T ---
\subsection{Evaluating the lattice temperature}\label{T}

To construct a true phase diagram we need to translate from the hopping parameter $\ka_s$
to the lattice temperature $T$. Therefore we need
to calculate the lattice temperature caused by particular choices of the $\ka_s$ hopping
parameter.

To do this we note that the presence of different hopping parameters in the temporal and spatial
directions means that correlation functions will also be anisotropic with respect to the temporal
and spatial axes of the lattice. The correlation length in lattice units in the time
direction, $\xi_t$, will be different from the space-like correlation length, $\xi_s$. However,
we expect the correlation lengths in physical units to be the same, even though the lattice
spacings $a_t$ and $a_s$ will be different. This means that $a_s \xi_s \equiv a_t \xi_t $, which
implies that
\beq{vary_T}
\frac{\xi_t}{\xi_s} = \frac{a_s}{a_t} = 2 T a_s = \frac{2 T}{\La_s}
\eeq
where $\La_s$ is the spatial cutoff. By varying the spatial hopping parameter, $\ka_s$, we will
vary the relative size of the two correlation lengths and so vary the effective lattice
anisotropy. This in turn allows us to vary the temperature \cite{MM}.

The overall scale of the lattice temperature is set by $N_t = 2$, which limits us to
high temperatures, but within this range of high values we can vary the temperature
{\it continuously} using the hopping parameter $\ka_s$.

Now (\ref{div_F}) gives

\setlength{\unitlength}{0.1cm}

\beqa{correlators}
\frac{\partial F}{\partial \ka_s} = - \frac{1}{N} \l\langle
\sum_{n, i}
\begin{picture}(10,1)
\put(1,1){\circle*{0.5}}
\put(1,1){\line(1,0){8}}
\put(9,1){\circle*{0.5}}
\end{picture}
\r\rangle &=&
- (d-1) \l\langle
\Phi^{n_4,\bf{n} *} \Phi^{n_4,\bf{n}+\bf{e}_i} + \Phi^{n_4,\bf{n}} \Phi^{n_4,\bf{n}+\bf{e}_i *}
\r\rangle = - (d-1) \xi_s \nonumber\\
\frac{\partial F}{\partial \ka_t}  = - \frac{1}{N} \l\langle
\sum_n
\begin{picture}(4,8)
\thicklines
\put(2,-3){\circle*{0.5}}
\put(2,-3){\line(0,1){8}}
\put(2,5){\circle*{0.5}}
\end{picture}
\r\rangle &=&
- \l\langle
(1+\mu_L)\Phi^{n_4,\bf{n} *} \Phi^{n_4+1,\bf{n}} + (1-\mu_L)\Phi^{n_4,\bf{n}} \Phi^{n_4+1,\bf{n} *}
\r\rangle =
- \xi_t \nonumber\\
&\Rightarrow& \frac{T}{\La_s} = \frac{(d-1)}{2}  \frac{\partial F}{\partial \ka_t}
\l/  \frac{\partial F}{\partial \ka_s} \r.
\eeqa

\begin{figure}[hp]
\begin{center}
\scalebox{1}{\includegraphics*[5.5cm,18.5cm][15cm,27cm]{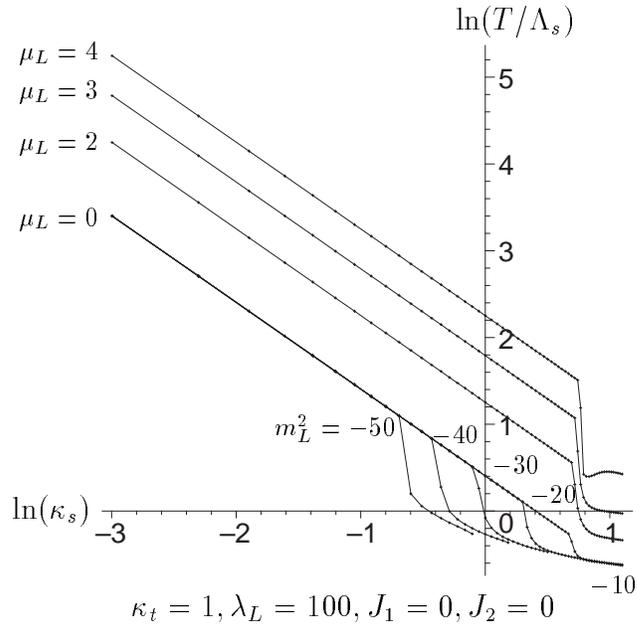}}
\caption{Effective T}
\label{f_T}
\end{center}
\end{figure}

Therefore purely by taking numerical derivatives with respect to the hopping parameters we can
construct the lattice temperature for a given value of $\ka_s$. As an example we
plot $\ln(T / \La_s)$ against $\ln(\ka_s)$ for a particular range of physical parameters in
Figure~\ref{f_T}.

In the unbroken regime of physical parameters there is a simple inverse relationship $\ka_s T = k$.
This is true for both zero and non-zero chemical potential. The only effect of the chemical
potential is to change the value of the constant $k$. In the unbroken regime we find an almost
perfect fit for the empirical curve

\beq{Ks_T}
\ka_s T = \frac{1}{2} \l( \mu_L^2 + 3 \r)
\eeq

However, as we pass through what is a second order phase transition, the relationship becomes
more complex. This break-off point occurs at lower and lower $\ka_s$ as one decreases
the lattice mass.

For any other set of physical parameters we can similarly evaluate the effective temperature
given by a particular choice of $\ka_s$.

% --- Phase transition using T ---
\subsection{Phase transition using $T$}\label{phase_T}

\begin{figure}[hp]
\begin{center}
\scalebox{1}{\includegraphics*[6.5cm,18.7cm][16cm,27.5cm]{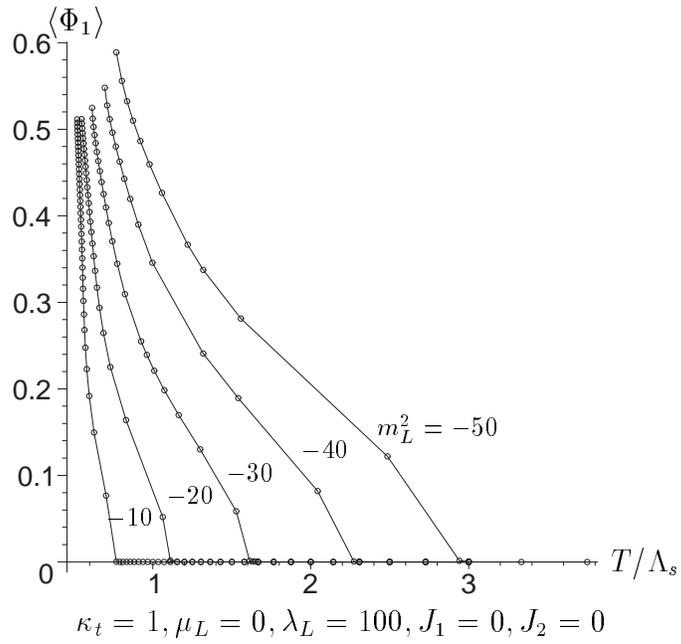}}
\caption{Phase transition with $T$}
\label{f_phi1T}
\end{center}
\end{figure}

We can now combine the information in sections \ref{phase_Ks}\ and \ref{T} to allow us to
evaluate $ \l\langle \Phi_1 \r\rangle $ as a function of $T$ and look for a phase transition.
This leads to the curves shown in Fig.~\ref{f_phi1T}.

These curves follow directly from Fig.~\ref{f_Ks} and, as expected, as the temperature is lowered
the symmetry breaks at some point, starting first with the lowest mass. The phase
transition is second order in form. As the temperature decreases towards zero there is an
unexpected divergence of the value of $\l\langle \Phi_1 \r\rangle$. However, this is precisely
in the regime where the approximation we are using is expected to break down, so no physical
interpretation should be made of this behaviour.

% --- Phase transition using m2 ---
\subsection{Phase transition using $m_L^2$}\label{phase_m2}

\begin{figure}[htb]
\begin{center}
\scalebox{1}{\includegraphics*[6.5cm,18.6cm][15.6cm,27.5cm]{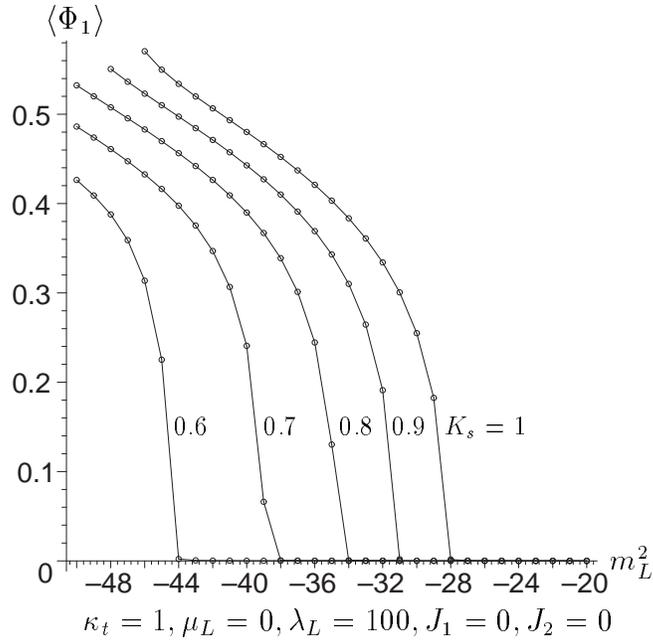}}
\caption{Phase transition with $m_L^2$}
\label{f_phi1m2}
\end{center}
\end{figure}

For completeness we show the standard symmetry breaking which occurs as one lowers the
physical mass. This second-order phase transition is discussed in much more detail
in \cite{EIM}. For particular choices of physical parameters we have the phase transition
curves shown in Figure~\ref{f_phi1m2}.

As expected, lowering the lattice mass breaks the symmetry, starting first with the
highest $\ka_s$ curve. The phase transition is second order.
The curves are performed at {\it fixed} hopping parameter values and
{\it not} at fixed lattice temperature. In the unbroken phase, as we can see in Fig.~\ref{f_T}, the
effective lattice temperature is {\it independent} of the lattice mass, so fixing $\ka_s$ is
equivalent to fixing $T$. However, when the symmetry is broken the temperature becomes mass
dependent. Although this make things more complex it does not stop us from measuring the
lattice temperature at the phase transition. Using the empirical formula (\ref{Ks_T}) we can
calculate the lattice temperatures. $\ka_s = 0.6, 0.7, 0.8, 0.9$ and $1.0$ correspond
to $T=2.499, 2.143, 1.875, 1.667$ and $1.5$ respectively.

% --- Phase transition using mu_L ---
\subsection{Phase transition using $\mu_L$}\label{phase_mu}

\begin{figure}[htb]
\begin{center}
\scalebox{1}{\includegraphics*[6.5cm,19.2cm][15.5cm,27.8cm]{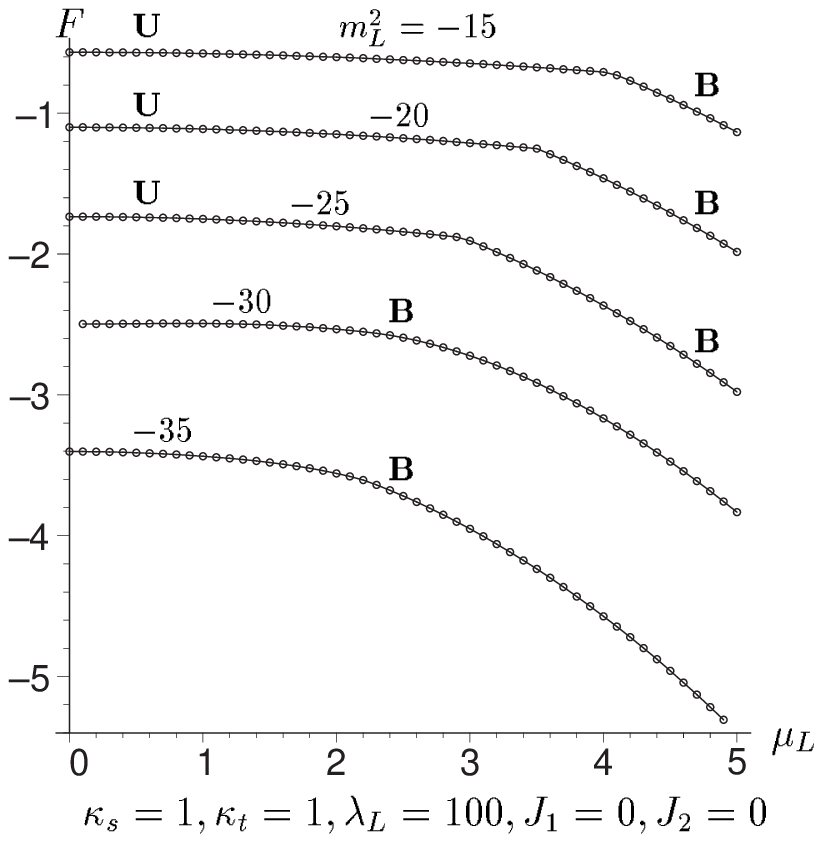}}
\caption{Phase transition in F with $\mu_L$}
\label{f_Fmu}
\end{center}
\end{figure}

Having seen how the symmetry is restored as one increases the temperature and lattice mass, we
should finally look at evaluating $ \l\langle \Phi_1 \r\rangle $ as a function of $\mu_L$. The
phase transition with $\mu_L$ involves travelling
between variational parameter spaces which look like Fig.~\ref{f_mini}(a) and Fig.~\ref{f_mini}(c).
This means that we will have first-order phase transitions. Examining the free energy across the
transition leads to the curves shown in Figure~\ref{f_Fmu}.

In the $m_L^2=-15 \cdots -25$ plots we can see that there is a kink in the free energy due to the
first order phase transition from an unbroken ({\bf \footnotesize U}) to a broken
({\bf \footnotesize B}) global $U(1)$ symmetry. The transition point moves to lower $\mu$ values
as the mass is decreases. These curves have variational parameter spaces which move between
Fig.~\ref{f_mini}(a) and Fig.~\ref{f_mini}(c). The kinks arise as one jumps from the unbroken to the
broken minima in the variational parameter space.

At $m_L^2 < -30$, and lower, the symmetry is broken for all $\mu$ values. The variational
parameter space moves between Fig.~\ref{f_mini}(b) and Fig.~\ref{f_mini}(d). The symmetry is already
broken at $\mu_L = 0$ because of the second order phase transition which occurs at low enough
temperatures, as discussed in section \ref{phase_m2}. The special case of the $m_L^2=-30$ curve is
discussed in more detail later.

\begin{figure}[htb]
\begin{center}
\scalebox{1}{\includegraphics*[6.5cm,19.2cm][15.3cm,27.3cm]{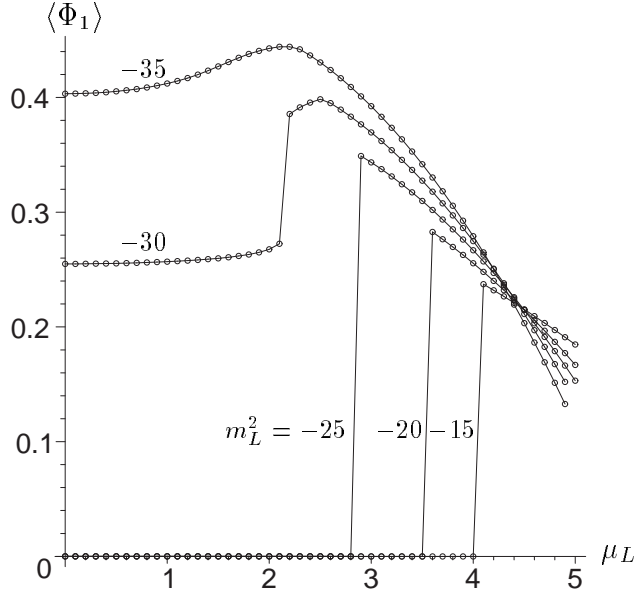}}
\caption{Phase transition in $\l \langle \Phi_1 \r \rangle$ with $\mu_L$}
\label{f_phi1mu}
\end{center}
\end{figure}

Using the free energy data, and again noting (\ref{phi1}), we can construct
the $ \l\langle \Phi_1 \r\rangle $ value. Where there is a kink in the $F$ value we expect
a first order phase transition in the field value. This is precisely the behaviour seen for the curves
in Figure~\ref{f_phi1mu}.

In the $m_L^2=-15 \cdots -25$ plots we see there is a first order
phase transition from an unbroken to a broken global $U(1)$ symmetry. Crossing through this phase
transition we move from Fig.~\ref{f_mini}(a) to Fig.~\ref{f_mini}(c) in variational space. Raising
the chemical potential breaks the symmetry, starting first with the lowest $m_L^2$ curve.
The transition point moves to higher $\mu_L$ values as one raises the lattice mass.

The $m_L^2 = -30$ case is more
unusual because, although the symmetry is already broken at $\mu_L=0$, we find evidence for a further
phase transition at $\mu_L=2.1$. It turns out that this unexpected behaviour occurs at the place
in phase space where the first order phase transition meets the second order phase transition.
We shall postpone discussion of this point, as it will become much clearer on consideration
of the full phase diagram, see section \ref{phase_2d}.

For $m_L^2<-30$ we see that the symmetry is already completely broken. There is a smooth crossover
as one increases the chemical potential value from zero and in doing this one
moves from Fig.~\ref{f_mini}(b) to Fig.~\ref{f_mini}(d) in variational parameter space.

Just as in section \ref{phase_m2} these curves are derived at {\it fixed} hopping parameter
values and {\it not} at fixed lattice temperature. However, unlike the lattice mass, the temperature
is chemical potential dependent, whether or not one is in the broken regime. Each point on the
curves is at a different lattice temperature. Again this make things more complex but we can still
find the lattice temperature at the phase transition.  We find that
that $\{m_L^2 = -15,\mu_L = 4\}$ gives $T=9.5$, $\{m_L^2 = -20,\mu_L = 3.5\}$ gives $T=7.625$,
 $\{m_L^2 = -25,\mu_L = 2.9\}$ gives $T=5.705$, and
finally $\{m_L^2 = -30,\mu_L = 2.1\}$ gives $T=3.705$.

% --- Phase diagram 2D ---
\subsection{Phase diagrams}\label{phase_2d}

\subsubsection{$\{T,\mu_L\}$ Phase diagram}

\begin{figure}[htb]
\begin{center}
\scalebox{1}{\includegraphics*[6.5cm,19.2cm][15.3cm,27.3cm]{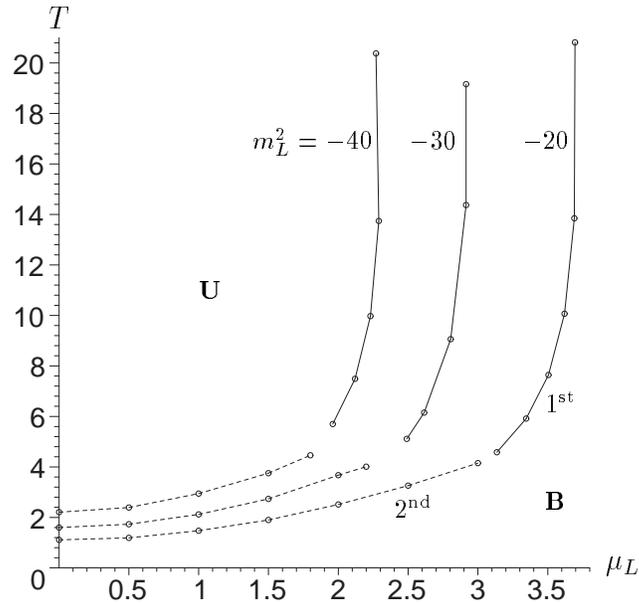}}
\caption{$\{T,\mu_L\}$ Phase diagram}
\label{f_Tmu}
\end{center}
\end{figure}

Combining all the information from the previous sections we can build up phase diagrams for the
theory in $\{T,\mu_L\}$ space. These are plotted for various value of the lattice mass in
Figure~\ref{f_Tmu}.

Close to the $\mu_L=0$ axis we have a second-order phase transition between the unbroken
phase at high temperatures and the broken phase at low temperatures. As one crosses through this
transition the minima plots change from Fig.~\ref{f_mini}(a) to Fig.~\ref{f_mini}(c). However, as
one follows this transition out to higher densities it becomes {\it first} order in nature. For
the first order phase transition the minima plots change from Fig.~\ref{f_mini}(a)
to Fig.~\ref{f_mini}(b).

\begin{figure}[htb]
\begin{center}
\scalebox{1}{\includegraphics*[6.5cm,17cm][15cm,27cm]{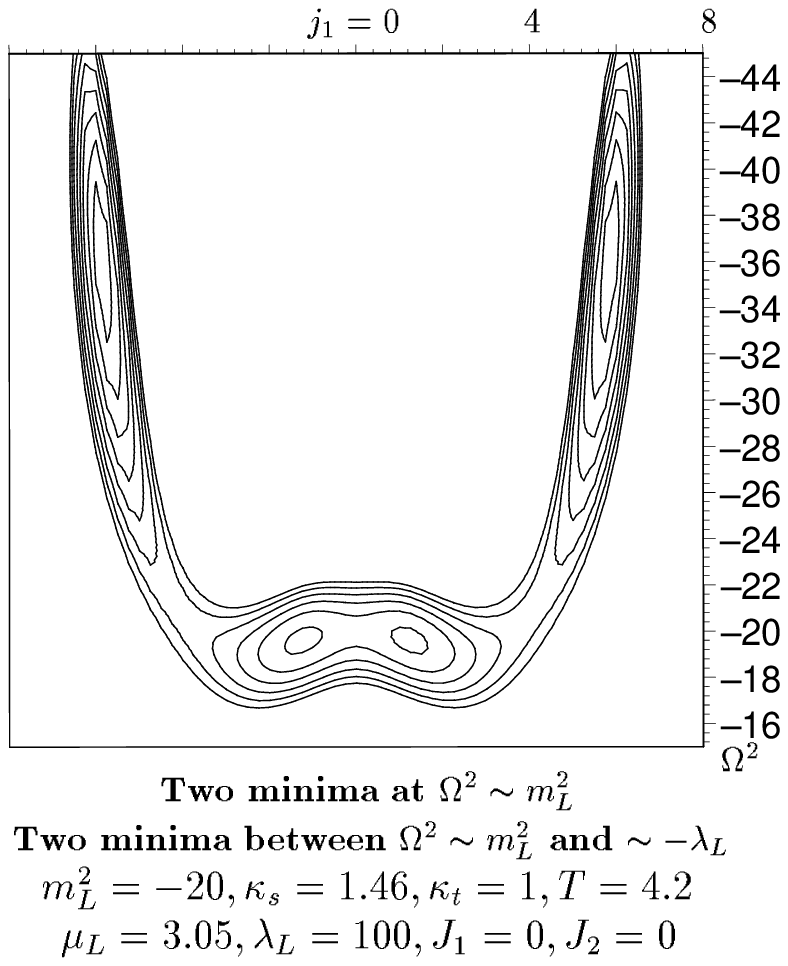}}
\caption{Changeover region}
\label{f_mini_co}
\end{center}
\end{figure}

For a small region at the changeover between the first-order and second-order phase transition
lines both the minima at $~-\la_L$ and the minima at $~m_L^2$ are present in variational parameter
space. Travelling from the second-order regime into the first-order part the minima at $~m_L^2$
spawn two further minima which rapidly move up towards $~-\la_L$. This is captured in the
variational parameter space picture seen in Figure~\ref{f_mini_co}.

As $\mu_L$ and $T$ are increased further one enters the first-order phase transition regime
and the two minima at $~m_L^2$ become one minimum.

It is this overlap which explains the unusual phase structure seen for the $m_L^2=-30$ curve
in Fig.~\ref{f_phi1mu}. We can that the unexpected additional first-order transition for
this curve sits at $\mu_L=2.1 , T = 3.7$, which is right in the changeover regime for
the $m_L^2=-30$ case in Fig.~\ref{f_Tmu}.
For $\mu_L < 2.1$ the field value is being evaluated at one of a pair of minima at $~m_L^2$, for
the $\mu_L > 2.1$ the field value is being evaluated at a pair of minima at $~-\la_L$. The jump
between the two causes the first-order phase transition.

\begin{figure}[htb]
\begin{center}
\scalebox{1}{\includegraphics*[5cm,23.7cm][17cm,27cm]{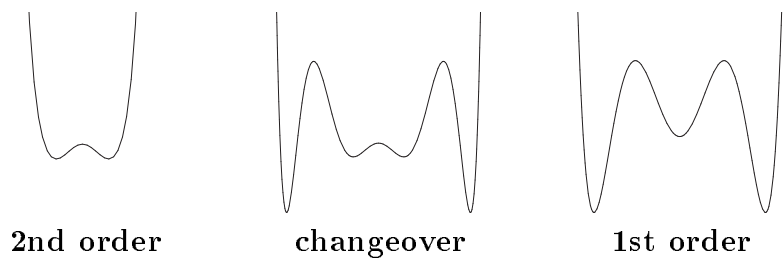}}
\caption{Qualitative $F$ against $\l\langle \Phi_1 \r\rangle$ curves along the phase 
transition line}
\label{curves}
\end{center}
\end{figure}

In physical space we have the three distinct types of pictures on the {\it broken} side of the
transition, these are shown in Fig.~\ref{curves}.

As one moves through the changeover area, starting from the second-order area of the phase
diagram, the first-order style metastable states appear out beyond the second-order style
minima. Then as one moves into the first-order area the two second-order minima merge
to give the classic first-order field distribution.

\subsubsection{$\{T,\rho\}$ Phase diagram}

\begin{figure}[htb]
\begin{center}
\scalebox{1}{\includegraphics*[3.5cm,18.4cm][18cm,27.3cm]{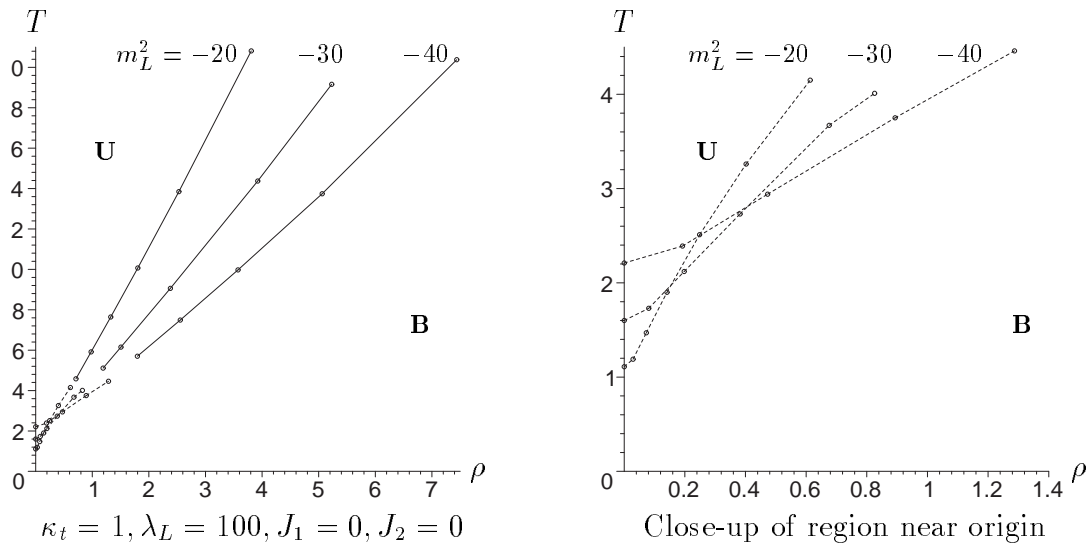}}
\caption{$\{T,\rho\}$ Phase diagram}
\label{f_TQ}
\end{center}
\end{figure}

One can also calculate the charge density in lattice units which, noting
that $\mathcal{H}_{eff} = \mathcal{H} + \mu \mathcal{Q}$ and $ \mu_L = a_t \mu $, is
\beq{charge}
\rho := \frac{\l\langle \mathcal{Q} \r\rangle}{N_s a_s^3}
= 2 T \frac{ \partial F}{ \partial \mu_L}
\eeq
where, because we are in lattice units, $a_s = \La_s = 1$. Using (\ref{charge}) one
can construct the phase diagram of temperature against charge density. This is plotted
in Figure~\ref{f_TQ}.

The lower ends of the transition lines move closer to the origin in $\{T,\rho\}$ space as one
increases the mass. In the close-up of the second-order phase transition lines, near to 
the $\mu=0$ axis, we can see an interesting complex curve form. As the transition 
becomes first-order for higher chemical potentials the transition lines become straighter.

% --- Conclusions ---
\section{Conclusions}\label{conclusions}

An LDE optimisation of the standard hopping parameter expansion has allowed access to some of
the truly non-perturbative physics of the scalar global $U(1)$ model. Close to the $\mu_L=0$ axis
we find the expected second-order phase transition but as we track out to higher chemical potentials
we see that this transition becomes {\it first} order. At our level of approximation at least there
appears to be evidence for a more complex phase structure than expected.

Although the $U(1)$ model is in itself intrinsically interesting as a simple model for Higgs
physics, the work in this paper can also be seen as a test of LDE techniques as applied to
phase transition at finite density in general. In this context we can see that the LDE is a
sucessful approach. The presence of the additional variational parameters allows one to examine a
theory as one tracks {\it through} the phase transition with any of the physical parameters.
By the choice of variational parameter one can actually choose which particular equivalent broken
vacuum state the model is in, as demonstrated in \cite{EIM}. Also the form of the free energy
contours in variational parameter space signals the phase-space behaviour of the physical model
itself. This allows one to extract the physical phase of the system without explicitly
calculating the particular order parameter such as $ \l\langle \Phi_1 \r\rangle$.

To extend our approach to plotting out a phase diagram at finite density to a
more complex model, in particular a gauge theory, we need to find an equivalent, lattice regularised,
expansion to optimise. The hopping parameter expansion is only available for scalar
theories: the derivative terms cannot be broken up in the same way for a gauge field theory.
However, there is an alternative lattice expansion in the strong coupling expansion, i.e. an expansion
in powers of the trace round lattice plaquettes. An example of an LDE optimisation of
a plaquette expansion for gauge theories on the lattice at zero temperature and chemical
potential is given in \cite{AJ93a}. Using a similar scheme to that used in this paper for scalars
one could extend considerations of gauge theories on the lattice to finite temperature
and densities.

% --- Acknowledgement ---
\section{Acknowledgements}

We are grateful to R.J.~Rivers for useful discussions. D.~Winder also gratefully acknowledges
the financial support of the Particle Physics and Astronomy Research Council.

% ******************************************************************

\renewcommand{\thesection}{\Alph{section}}
\setcounter{section}{0}

% --- Appendix A---
\section{Diagrams}\label{diagrams}

Below we list all the diagrams which contribute to the cumulant expansion for the free
energy of a $U(1)$ scalar Higgs model with finite temperature and chemical potential up
and including order 3. With a lattice of two temporal links in extent we
have periodicity across some of the diagrams. Points which are identified as identical
are signified with an open circle in the diagrams. In the multiplicities that accompany
the diagrams we have omitted the overall $N$ factor and used the shorthand
\beq{ds}
d_j = 2(d - 1) - j,
\eeq
where $d$ is the spacetime dimension of the lattice and $j \in \mathcal{Z}$.

\setlength{\unitlength}{0.10cm}
\setlength{\textheight}{24.0cm} % 25cm for A4, 23cm for Letter or DJ

\begin{tabular}{{c}{c}{c}{c}}
%
% ORDER 1 DIAGRAMS
%
\begin{picture}(12,12)
\thinlines
\put(2,1){\circle*{0.5}}
\put(2,1){\line(1,0){8}}
\put(10,1){\circle*{0.5}}
\end{picture}
& $ m_{1,1} = \frac{1}{2} d_0$  &
\begin{picture}(5,12)
\thicklines
\put(3,2){\circle*{0.5}}
\put(3,2){\vector(0,1){8}}
\put(3,10){\circle*{0.5}}
\end{picture}
& $ m_{1,2} = 1$ \\
%
% ORDER 2 DIAGRAMS
% 0 temporal links
%
\begin{picture}(12,12)
\thinlines
\put(2,2){\circle*{0.5}}
\put(2,1){\line(1,0){8}}
\put(2,3){\line(1,0){8}}
\put(10,2){\circle*{0.5}}
\end{picture}
& $m_{2,1} = \frac{1}{2} d_0$  &
\begin{picture}(24,12)
\thinlines
\put(6,2){\circle*{0.5}}
\put(6,2){\line(1,0){8}}
\put(14,2){\circle*{0.5}}
\put(14,2){\line(1,0){8}}
\put(22,2){\circle*{0.5}}
\end{picture}
&$ m_{2,2} = d_0 d_1 $\\
%
% ORDER 2 DIAGRAMS
% 1 temporal link
%
\begin{picture}(12,12)
\thicklines
\put(2,0){\circle*{0.5}}
\put(2,0){\vector(0,1){8}}
\put(2,8){\circle*{0.5}}
\thinlines
\put(2,0){\line(1,0){8}}
\put(10,0){\circle*{0.5}}
\end{picture}
&$ m_{2,3} = 2 d_0 $ &
\begin{picture}(12,12)
\thicklines
\put(2,0){\circle*{0.5}}
\put(2,0){\vector(0,1){8}}
\put(2,8){\circle*{0.5}}
\thinlines
\put(2,8){\line(1,0){8}}
\put(10,8){\circle*{0.5}}
\end{picture}
&$ m_{2,4} = 2 d_0$ \\
%
% ORDER 2 DIAGRAMS
% 2 temporal links
%
\begin{picture}(5,12)
\thicklines
\put(3,0){\circle*{0.5}}
\put(2,0){\vector(0,1){8}}
\put(4,0){\vector(0,1){8}}
\put(3,8){\circle*{0.5}}
\end{picture}
&$ m_{2,5} = 1$  &
\begin{picture}(5,24)
\thicklines
\put(3,0){\circle{1}}
\put(3,0){\vector(0,1){8}}
\put(3,8){\circle*{0.5}}
\put(3,8){\vector(0,1){8}}
\put(3,16){\circle{1}}
\end{picture}
&$ m_{2,6} = 2$ \\
%
% ORDER 3 DIAGRAMS
% 0 temporal links
%
\begin{picture}(12,12)
\thinlines
\put(2,2){\circle*{0.5}}
\put(2,1){\line(1,0){8}}
\put(2,2){\line(1,0){8}}
\put(2,3){\line(1,0){8}}
\put(10,2){\circle*{0.5}}
\end{picture}
&$ m_{3,1} = \frac{1}{2} d_0 $ &
\begin{picture}(24,12)
\thinlines
\put(6,2){\circle*{0.5}}
\put(6,1){\line(1,0){8}}
\put(6,3){\line(1,0){8}}
\put(14,2){\circle*{0.5}}
\put(14,2){\line(1,0){8}}
\put(22,2){\circle*{0.5}}
\end{picture}
&$ m_{3,2} = 3 d_0 d_1$ \\
\begin{picture}(36,12)
\thinlines
\put(10,1){\circle*{0.5}}
\put(10,1){\line(1,0){8}}
\put(18,1){\circle*{0.5}}
\put(18,1){\line(1,0){8}}
\put(26,1){\circle*{0.5}}
\put(26,1){\line(1,0){8}}
\put(34,1){\circle*{0.5}}
\end{picture}
&$ m_{3,3} = 3 d_0 d_1^2 $ &
\begin{picture}(24,12)
\thinlines
\put(6,1){\circle*{0.5}}
\put(6,1){\line(1,0){8}}
\put(14,1){\circle*{0.5}}
\put(14,1){\line(4,3){8}}
\put(22,7){\circle*{0.5}}
\put(14,1){\line(1,0){8}}
\put(22,1){\circle*{0.5}}
\end{picture}
&$ m_{3,4} = d_0 d_1 d_2 $\\
%
% ORDER 3 DIAGRAMS
% 1 temporal link
%
\begin{picture}(12,12)
\thicklines
\put(2,1){\circle*{0.5}}
\put(2,1){\vector(0,1){8}}
\thinlines
\put(2,1){\circle*{0.5}}
\put(2,0){\line(1,0){8}}
\put(2,2){\line(1,0){8}}
\put(10,1){\circle*{0.5}}
\end{picture}
&$ m_{3,5} = 3 d_0 $ &
\begin{picture}(12,12)
\thicklines
\put(2,0){\circle*{0.5}}
\put(2,0){\vector(0,1){8}}
\thinlines
\put(2,8){\circle*{0.5}}
\put(2,7){\line(1,0){8}}
\put(2,9){\line(1,0){8}}
\put(10,8){\circle*{0.5}}
\end{picture}
&$ m_{3,6} = 3 d_0$ \nonumber \\
\begin{picture}(24,12)
\thicklines
\put(6,0){\circle*{0.5}}
\put(6,0){\vector(0,1){8}}
\thinlines
\put(6,0){\circle*{0.5}}
\put(6,0){\line(1,0){8}}
\put(14,0){\circle*{0.5}}
\put(14,0){\line(1,0){8}}
\put(22,0){\circle*{0.5}}
\end{picture}
&$ m_{3,7} = 6 d_0 d_1 $ &
\begin{picture}(24,12)
\thicklines
\put(14,0){\circle*{0.5}}
\put(14,0){\vector(0,1){8}}
\put(14,8){\circle*{0.5}}
\thinlines
\put(6,0){\circle*{0.5}}
\put(6,0){\line(1,0){8}}
\put(14,0){\circle*{0.5}}
\put(14,0){\line(1,0){8}}
\put(22,0){\circle*{0.5}}
\end{picture}
&$ m_{3,8} = 3 d_0 d_1$ \nonumber \\
\begin{picture}(12,12)
\thicklines
\put(2,0){\circle*{0.5}}
\put(2,0){\vector(0,1){8}}
\thinlines
\put(2,0){\circle*{0.5}}
\put(2,0){\line(1,0){8}}
\put(10,0){\circle*{0.5}}
\put(2,8){\line(1,0){8}}
\put(10,8){\circle*{0.5}}
\end{picture}
&$ m_{3,9} = 6 d_0^2 $ &
\begin{picture}(24,12)
\thicklines
\put(14,0){\circle*{0.5}}
\put(14,0){\vector(0,1){8}}
\thinlines
\put(6,8){\circle*{0.5}}
\put(6,8){\line(1,0){8}}
\put(14,8){\circle*{0.5}}
\put(14,8){\line(1,0){8}}
\put(22,8){\circle*{0.5}}
\end{picture}
&$ m_{3,10} = 3 d_0 d_1$ \nonumber \\
\begin{picture}(24,12)
\thicklines
\put(6,0){\circle*{0.5}}
\put(6,0){\vector(0,1){8}}
\thinlines
\put(6,8){\circle*{0.5}}
\put(6,8){\line(1,0){8}}
\put(14,8){\circle*{0.5}}
\put(14,8){\line(1,0){8}}
\put(22,8){\circle*{0.5}}
\end{picture}
& $m_{3,11} = 6 d_0 d_1$  &
%
% ORDER 3 DIAGRAMS
% 2 temporal links
%
\begin{picture}(12,24)
\thicklines
\put(2,0){\circle{1}}
\put(2,0){\vector(0,1){8}}
\put(2,8){\circle*{0.5}}
\put(2,8){\vector(0,1){8}}
\put(2,16){\circle{1}}
\thinlines
\put(2,0){\line(1,0){8}}
\put(10,0){\circle*{0.5}}
\end{picture}
& $m_{3,12} = 3 d_0$ \nonumber \\
\begin{picture}(12,12)
\thicklines
\put(2,0){\circle*{0.5}}
\put(1,0){\vector(0,1){8}}
\put(3,0){\vector(0,1){8}}
\put(2,8){\circle*{0.5}}
\thinlines
\put(2,0){\line(1,0){8}}
\put(10,0){\circle*{0.5}}
\end{picture}
& $m_{3,13} = 3 d_0 $ &
\begin{picture}(12,12)
\thicklines
\put(2,0){\circle*{0.5}}
\put(2,0){\vector(0,1){8}}
\put(10,0){\vector(0,1){8}}
\put(2,8){\circle*{0.5}}
\thinlines
\put(2,0){\line(1,0){8}}
\put(10,0){\circle*{0.5}}
\end{picture}
& $m_{3,14} = 3 d_0$ \nonumber \\
\begin{picture}(12,24)
\thicklines
\put(2,10){\circle*{0.5}}
\put(2,10){\vector(0,1){8}}
\put(2,10){\circle*{0.5}}
\put(10,2){\vector(0,1){8}}
\thinlines
\put(2,10){\line(1,0){8}}
\put(10,10){\circle*{0.5}}
\end{picture}
& $m_{3,15} = 6 d_0 $ &
\begin{picture}(12,12)
\thicklines
\put(2,0){\circle*{0.5}}
\put(1,0){\vector(0,1){8}}
\put(3,0){\vector(0,1){8}}
\put(2,8){\circle*{0.5}}
\thinlines
\put(2,8){\line(1,0){8}}
\put(10,8){\circle*{0.5}}
\end{picture}
& $m_{3,16} = 3 d_0$ \nonumber \\
\begin{picture}(12,12)
\thicklines
\put(2,0){\circle*{0.5}}
\put(2,0){\vector(0,1){8}}
\put(10,0){\circle*{0.5}}
\put(10,0){\vector(0,1){8}}
\thinlines
\put(2,8){\circle*{0.5}}
\put(2,8){\line(1,0){8}}
\put(10,8){\circle*{0.5}}
\end{picture}
& $m_{3,17} = 3 d_0$  &
%
% ORDER 3 DIAGRAMS
% 3 temporal links
%
\begin{picture}(5,12)
\thicklines
\put(3,0){\circle*{0.5}}
\put(2,0){\vector(0,1){8}}
\put(3,0){\vector(0,1){8}}
\put(4,0){\vector(0,1){8}}
\put(3,8){\circle*{0.5}}
\end{picture}
& $m_{3,18} = 1$ \\
\begin{picture}(5,24)
\thicklines
\put(3,0){\circle{1}}
\put(2,0){\vector(0,1){8}}
\put(4,0){\vector(0,1){8}}
\put(3,8){\circle*{0.5}}
\put(3,8){\vector(0,1){8}}
\put(3,16){\circle{1}}
\end{picture}  & $m_{3,19} = 3$  &  \nonumber

\end{tabular}


\begin{thebibliography}{99}

\bibitem{MM}
I.~Montvay and G.~M\"unster
\ttitle{Quantum Fields on a Lattice}
(Cambridge Monographs on Mathematical Physics, Cambridge, 1994).

\bibitem{K}
J.I.~Kapusta,
\ttitle{Finite Temperature Field Theory}
(Cambridge University Press, Cambridge, 1989)

\bibitem{Y}
J.M.~Yeomans,
\ttitle{Statistical Mechanics of Phase Transitions}
(Oxford Science Publications, Oxford, 1992)

%%%%%%%%%%%%%%%%%  LDE AND VARIOUS VARIANTS THEREOF %%%%%%%%%%%%%%%


\bibitem{EIM}
T.S.~Evans, M.~Ivin and M.~M\"obius,
\arttitle{An optimized pertubation expansion for a global O(2) theory}
Nucl. Phys.\vol{B577} (2000) 325.
% Tim+MMs lattice O(2) paper




\bibitem{Jo}
H.F.~Jones,
Nucl. Phys. (Proc.Suppl.) \vol{B39} (1995) 220.

\bibitem{JP}
H.F.~Jones and P.~Parkin,
\arttitle{The Renormalised Thermal Mass with Non-zero charge density}
[\eprint{hep-th/0005069}].
% Phils continuum T, mu O(2) paper

\bibitem{JPW}
H.F.~Jones, P.~Parkin and D.~Winder,
\arttitle{Quantum Dynamics of the Slow Rollover Transition in the
Linear Delta Expansion}
[\eprint{hep-th/0008069}].
% Slowroll LDE QM paper

\bibitem{AJ93a}
J.O.~Akeyo and H.F.~Jones,
Phys. Rev. D \vol{47} (1993) 1668.
% SU(2) lattice PMS

\bibitem{AJ93b}
J.O.~Akeyo and H.F.~Jones,
Z. Phys. C \vol{58} (1993) 629.
% SU(2)-SO(3) lattice PMS

\bibitem{AJP}
J.O.~Akeyo, H.F.~Jones and C.S.~Parker,
\arttitle{Extended Variational Approach to the SU(2) Mass Gap on the
Lattice}
Phys. Rev. D \vol{51} (1995) 1298\tpre{
[\eprint{hep-ph/9405311}]}.
% SU(2) lattice PMS

\bibitem{EJR98a}
T.S.~Evans, H.F.~Jones and A.~Ritz,
\arttitle{An Analytical Approach to Lattice Gauge-Higgs Models}
in \inproctitle{Strong and Electroweak Matter '97},
ed. F.Csikor and Z.Fodor (World Scientific, Singapore,
1998\tpre{, ISBN {\tt 981-02-3257-8}})\tpre{
[\eprint{hep-ph/9707539}]}.

\bibitem{EJR98b}
T.S.~Evans, H.F.~Jones and A.~Ritz,
\arttitle{On the Phase Structure of the 3D SU(2)-Higgs Model and the
Electroweak Phase Transition}
Nucl. Phys. \vol{B517} (1998) 599\tpre{
[\eprint{hep-ph/9710271}]}.

\bibitem{DJ}
A.~Duncan and H.F.~Jones,
Phys. Rev. D \vol{47} (1993) 2560.
% QM LDE

\bibitem{BuDJ}
I.R.C.~Buckley, A.~Duncan and H.F.~Jones,
Phys. Rev. D \vol{47} (1993) 2554.
% 0 Dim. LDE convergence of Z

\bibitem{BeDJ}
C.M.~Bender, A.~Duncan and H.F.~Jones,
Phys. Rev. D \vol{49} (1994) 4219.
% 0 Dim. LDE convergence of ln(Z)




\bibitem{BMPS}
C.M.~Bender, K.A.~Milton, S.S.~Pinsky and L.M.~Simmons Jr.,
J. Math. Phys. \vol{30} (1989) 1447.
% Non Linear delta expansion, ODEs



\bibitem{KM}
W.~Kerler and T.~Metz,
Phys. Rev. D \vol{44} (1991) 1263.
% Pure gauge SU(2), FAC



\bibitem{ZTW}
X.-T.~Zheng, Z.G.~Tan and J.~Wang,
Nucl. Phys. \vol{B287} (1987) 171.
% Pure Gauge SU(2), MFE

\bibitem{ZL}
X.-T.~Zheng, B.S.~Liu,
Intl. J. Mod. Phys. A \vol{6} (1991) 103.
% SU(2)-Higgs, MFE, Adam found this was actually doing very odd things.



\bibitem{ABS}
Jens O.~Andersen, Eric~Braaten and Michael~Strickland, The massive thermal
basketball diagram\tpre{[\eprint{ hep-ph/0002048 }]}.
% Screened perturbation theory paper



\bibitem{SSS}
A.N.~Sissakian, I.L.~Solovtsov and O.P.~Solovtsova,
Phys. Lett. B \vol{321} (1994) 381.



\bibitem{Yu}
V.I.~Yukalov,
J. Math. Phys. \vol{33} (1992) 3994.
% Mixed PMS and Minimal Difference



\bibitem{Wu}
C.M.~Wu et al.,
\arttitle{Phase Structure of Lattice $\phi^4$ Theory by
Variational Cumulant Expansion}
Phys. Lett. B \vol{216} (1989) 381.
% Lattice real phi**4 M1FE



\bibitem{Ca}
W.E.~Caswell,
Ann. Phys. \vol{123} (1979) 153.
% QM LDE



\bibitem{Ki}
J.~Killingbeck,
J. Phys. A \vol{14} (1981) 1005.
% QM LDE



\bibitem{Ok}
A.~Okapi\'{n}ska,
Phys. Rev. D \vol{35} (1987) 1835.
% LDE, QFT in continuum, also QM and 0 dim.
% See also \cite{Bu} section 1.3.4 for Gross-Neveu, NJL etc.

\bibitem{Ok2}
A.~Okapi\'{n}ska,
Phys. Rev. D \vol{36} (1987) 2415.
%"Nonstandard Expansion Techniques for the Finite Temperature Effective
%Potential in $\lambda\Phi^{4}$ Quantum Field Theory",



\bibitem{Ya}
J.M.~Yang,
J. Phys. G \vol{17} (1991) L143.
% U(1) Higgs, Fixed Higgs modulus, M1FE, no higgs mass varm. param.

\bibitem{YWZ}
J.M.~Yang, C.M.~Wu and P.Y.~Zhao,
J. Phys. G \vol{18} (1992) L1.
% U(1) Higgs, variable Higgs modulus, M1FE, higgs mass varm. param.




%%%%%%%%%%%%%%%%%  STEVENSON AND PMS %%%%%%%%%%%%%%%


\bibitem{PMS}
P.M.~Stevenson,
Phys. Rev. D \vol{23} (1981) 2916.

\bibitem{PDB}
D.E.~Groom et al,
Particle Physics Databook,
Eur. Phys. J. \vol{C15} 1 (2000)


%%%%%%%%%%%%%%%%%  MU=/= 0 FORMALISMS %%%%%%%%%%%%%%%

\bibitem{E} {T.S.~Evans}, {\em in}
{``Fourth Workshop on Thermal Field Theories and their
Applications''}, 5-10 August 1995, Dalian, China, ed. Y.X.~Gui,
F.C.~Khanna, and Z.B.~Su (World Scientific, Singapore, 1996)
p.283-295 ({hep-ph/9510298}).

%%%%%%%%%%%%%%%%%  LATTICE RESULT AND LCE %%%%%%%%%%%%%%%



\bibitem{LW1} M.~L\"{u}scher and P.~Weisz, Application of the Linked Cluster
expansion to the n-component $\phi^4$ theory (Nuclear Physics B {\bf 300}, 325-359, 1988)

\bibitem{LW2}
M.~L\"uscher and P.~Weisz,
Nucl. Phys. \vol{B300} (1989) 325.
% O(N), LCE



\bibitem{Wo}
M.~Wortis,
\arttitle{Linked Cluster Expansion} in \inproctitle{Phase
Transitions and Critical Phenomena}, vol.3, eds. C.~Domb and
M.S.~Green (Academic Press, London, 1974).
% Linked Cluster Expansion.



\bibitem{R1} T.~Reisz, Hopping Parameter Series construction for Models with
Nontrivial Vacuum (hep-lat/9802023, 17 Feb 1998)

\bibitem{R2} T.~Reisz, Advanced Linked Cluster Expansion. Scalar fields at finite
temperature (hep-lat/9505023, 29 May 1995)

\bibitem{HR}
H.~Meyer-Ortmanns and T.~Reisz,
Nucl. Phys. (Proc.Suppl.) \vol{B73} (1999) 892
% LATTICE98 talk, automatic LCE, SU(2) Higgs




\end{thebibliography}
\end{document}